\documentclass[%
 reprint,
 superscriptaddress,
 amsmath,amssymb,
 aps,
prl
]{revtex4-2}

\usepackage{graphicx}
\usepackage{dcolumn}
\usepackage{bm}
\usepackage{hyperref}
\usepackage{url}
\hypersetup{hidelinks}
\usepackage[dvipsnames]{xcolor}
\usepackage{makecell}
\usepackage{physics}
\usepackage{upgreek}
\usepackage{float}
\usepackage{amsmath}
\usepackage{siunitx}
\usepackage{bbding}
\usepackage{multirow}
\usepackage{mathrsfs}
\usepackage{gensymb}
\usepackage{float}

\begin{document}

\title{Spin Orientation Driven Polarization in Collinear Magnets}

\author{Yixun Zhang}
\altaffiliation{These authors contributed equally to this work}
\affiliation{Key Laboratory of Material Simulation Methods and Software of Ministry of Education, College of Physics, Jilin University, Changchun 130012, China}

\author{Longju Yu}
\altaffiliation{These authors contributed equally to this work}
\affiliation{Key Laboratory of Material Simulation Methods and Software of Ministry of Education, College of Physics, Jilin University, Changchun 130012, China}

\author{Yizhou Tong}
\affiliation{College of Physics, Jilin University, Changchun 130012, China}

\author{Ying Sun}
\affiliation{Key Laboratory of Material Simulation Methods and Software of Ministry of Education, College of Physics, Jilin University, Changchun 130012, China}

\author{Xu Li}
\affiliation{Department of Physics, Hunan Institute of Advanced Sensing and Information Technology, Xiangtan University, Xiangtan 411105, China}

\author{Hong Jian Zhao}
\affiliation{Key Laboratory of Material Simulation Methods and Software of Ministry of Education, College of Physics, Jilin University, Changchun 130012, China}
\affiliation{International Center of Future Science, Jilin University, Changchun 130012, China}
\affiliation{Institute of Quantum Science and Technology, Yanbian University, Yanji 133002, China}

 \author{Yanming Ma}
\affiliation{School of Physics, Zhejiang University, Hangzhou 310058, China}
 \affiliation{Key Laboratory of Material Simulation Methods and Software of Ministry of Education, College of Physics, Jilin University, Changchun 130012, China}

\date{\today}

\begin{abstract}
In a collinear magnet, the predominant magnetic moments are collectively aligned along a specific spatial orientation, and this alignment may yield intriguing phenomena such as spin orientation driven polarization. It is well known that spin orientation driven polarization is a relativistic effect that widely occurs in various type-II multiferroics. However, a universal theory that describes such a phenomenon and directs the corresponding materials discovery is lacking. Here, we revisit the magnetic structures of collinear magnets and explore the spin-orientation-dependent phenomena therein. Based on symmetry principles, we analyze the spin point groups (SPGs) that are associated with collinear magnets in the non-relativistic regime, demonstrate how relativistic spin-orbit interaction reduces each SPG to various magnetic point groups that are associated with different magnetic alignments, and classify the SPGs with respect to spin orientation driven polarization. We employ our theory to elucidate the mechanisms of spin orientation driven polarization in a variety of type-II multiferroics. Combined with first-principles simulations, we further show that polarization may be driven in nonpolar collinear antiferromagnets (e.g., CuFeS$_2$) by reorienting their magnetic alignments. Our theory provides guidelines for designing and discovering materials with spin orientation driven polarization, which will benefit the development of spintronics based on type-II multiferroics and related materials.
\end{abstract}

\maketitle

\noindent
\textit{Introduction.} Magnetic materials with strongly coupled polar and magnetic orders are not only of fundamental interest in condensed matter physics but also of technological importance for developing high-performance information devices~\cite{2006multiferroic,magnetoelectri,MagnetoelectricDevices,spaldin2019advances,Dong02112015,tokura2014multiferroics,fiebig2016evolution,Arimareview}. Type-II multiferroics are such materials~\cite{tokura2014multiferroics,Dong02112015,fiebig2016evolution,Arimareview} whose polarizations are driven by and coupled with magnetic orders via exchange striction~\cite{fiebig2016evolution,Arimareview,HoMnO3exchangestr}, inverse Dzyaloshinskii-Moriya interaction~\cite{fiebig2016evolution,Arimareview,KNBspincurrent,DMI}, or spin-orientation-dependent mechanism~\cite{fiebig2016evolution,Arimareview,spinbond,Arimapd,LMCOHJX,LMCOHJZ}. The third mechanism is rooted in the relativistic spin-orbit interaction (SOI) and is responsible for the spin orientation driven polarization in a large variety of collinear magnets with simple magnetic structures (see e.g., Refs.~\cite{BCGEPicozzi,BCGOpolarline,BCGEdiffmagstr,Ba2CoGe2O7,Ba2CoGe2O72,Ba2MnGe2O7,Ba2MnGe2O7prb,Sr2CoSi2O7,SCSOmag,Sr2CoMnGeO,SrCoGeOpolar,CCSOapljapanmag,CCSOnc,Ba2FeSi2O7,BFSO2,LaMnCrOlong,LMCOHJZ,LMCOHJX,TbMnCrO,TbMnCrOtheory,SmMnCrO,Cu2OSeO3science,Cu2OSeO3prb,Cu2OSeO3,MgFeN,OsX2,VSVSe,VSeTeRuBr,GdI2}). It is found that the magnetic moments in such materials are aligned in a collinear manner or slightly noncollinear manner, and the orientations of the predominant magnetic moments determine their polarizations~\cite{tokura2014multiferroics,fiebig2016evolution,Arimareview}. Such a phenomenon has been experimentally observed in various type-II multiferroics~\cite{Ba2CoGe2O7,Ba2CoGe2O72,Ba2MnGe2O7,Ba2MnGe2O7prb,Sr2CoSi2O7,SCSOmag,Sr2CoMnGeO,SrCoGeOpolar,CCSOapljapanmag,CCSOnc,Ba2FeSi2O7,BFSO2,LaMnCrOlong,TbMnCrO,SmMnCrO,Cu2OSeO3science,Cu2OSeO3prb} such as Ba$_2$CoGe$_2$O$_7$~\cite{Ba2CoGe2O7,Ba2CoGe2O72,Ba2MnGe2O7,BCGEdiffmagstr,BCGEPicozzi,BCGOpolarline}, LaMn$_3$Cr$_4$O$_{12}$~\cite{LaMnCrOlong,LMCOHJZ,LMCOHJX} and Cu$_2$OSeO$_3$~\cite{Cu2OSeO3science,Cu2OSeO3prb,Cu2OSeO3}, and computationally predicted in MgFe$_2$N$_2$~\cite{MgFeN}, OsI$_2$~\cite{OsX2} and other~\cite{VSVSe,VSeTeRuBr,GdI2} type-II multiferroics. Furthermore, Ref.~\cite{CaFeOpicozzi} predicts that the polarization in CaFeO$_2$ and MgFeO$_2$ collinear antiferromagnets is controllable by reorienting the directions of their predominant magnetic moments. Apparently, the aforementioned phenomenon is associated with the symmetry breakings caused by the magnetic alignments~\cite{tokura2014multiferroics,fiebig2016evolution,Arimareview}; yet, a universal theory that describes spin orientation driven polarization is lacking. This hinders the understanding of spin-orientation-dependent polarization in magnets and the discovery of such a phenomenon in a broad range of magnetic materials.

\begin{figure*}[hbtp]
\includegraphics[width=1.0\linewidth]{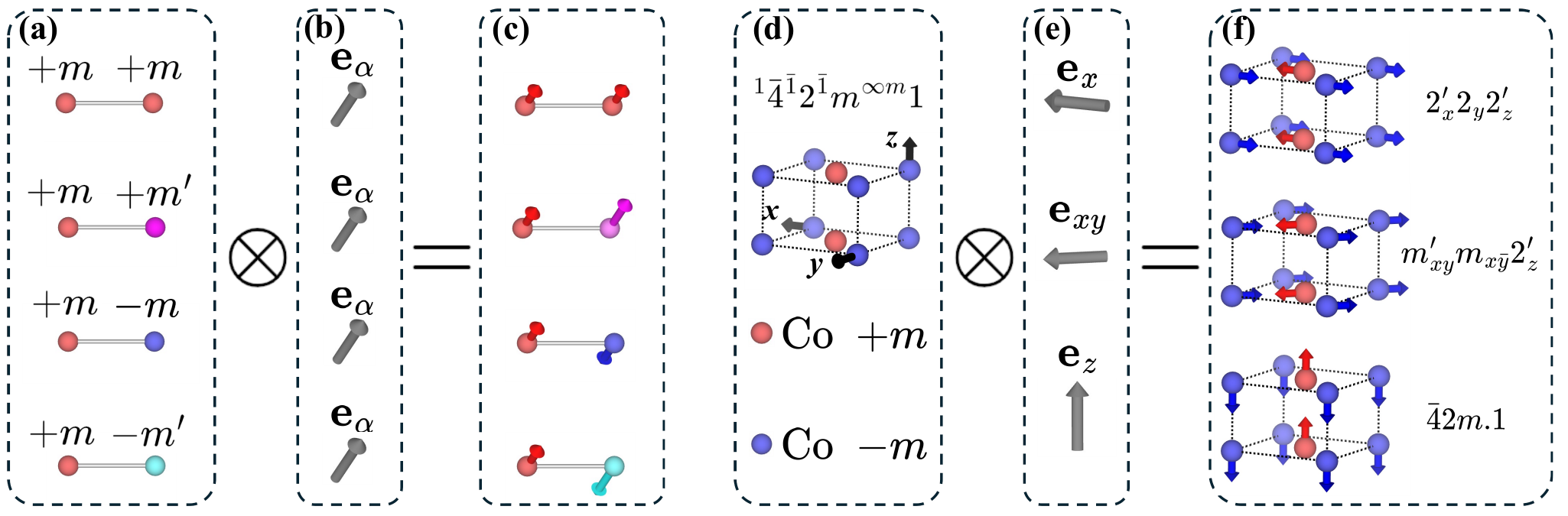} 
\caption{Examples of scalar and vector magnetic structures. Panel (a) sketches four different scalar magnetic structures. The $+m$, $+m^\prime$, $-m$ and $-m^\prime$ scalar magnetic moments are represented by red, pink, blue and cyan spheres, respectively. Panel (b) sketches the $\mathbf{e}_\alpha$ orientation vector that describes the magnetic alignment along the $\alpha$ direction ($\mathbf{e}_\alpha$ being a unit vector). Panel (c) shows four vector magnetic structures resulting from the direct products between the scalar magnetic structures in panel (a) and  the $\mathbf{e}_\alpha$ orientation vector in panel (b). Panel (d) is a scalar magnetic structure for Ba$_2$CoGe$_2$O$_7$. Panel (e) sketches $\mathbf{e}_x$, $\mathbf{e}_{xy}$ and $\mathbf{e}_z$ orientation vectors. Panel (f) shows three vector magnetic structures resulting from the direct products between the scalar magnetic structure in panel (d) and the orientation vectors in panel (e). In panels (d)--(f), Ba, Ge, and O ions are not shown.}
\label{fig:reorientation}
\end{figure*}

In the present work, we explore the spin-orientation-dependent phenomena in collinear magnets by group-theoretical techniques. We demonstrate that a collinear magnetic structure is a direct product between a scalar magnetic structure and an orientation vector characterizing the magnetically aligned direction. According to spin group theory~\cite{Libor2022Beyond,Libor2022Emerging,lqh2022}, the symmetry of a collinear magnetic structure can be described by a spin point group (SPG) in the non-relativistic regime; the SPG symmetry is reduced to the magnetic point group (MPG) symmetry when considering the relativistic SOI. We show that an SPG can be mapped to various MPGs associated with different orientation vectors. Focusing on collinear magnets, we develop a universal theory that describes the correlations among SPG, MPG and the orientation vector, and provide a symmetry classification of SPGs with respect to spin orientation driven polarization. Our theory explains the origin of polarization in a variety of type-II multiferroics and suggests the possibility of creating polarization in  nonpolar collinear magnets by spin reorientation. The latter is supported by our first-principles simulations which show that nonpolar CuFeS$_2$ exhibits spin-orientation-dependent polarization.\\

\noindent
\textit{Spin orientation and symmetry.} We focus on collinear magnets whose magnetic moments are aligned collinearly. In case of spin canting, the magnetic structure might be slightly noncollinear but can still be seen as collinear when working with its predominant magnetic moments. A collinear magnetic structure is referred to as a vector magnetic structure, which is a direct product between a scalar magnetic structure and an orientation vector characterizing the magnetic alignment. Figure~\ref{fig:reorientation}(a) demonstrates four types of scalar magnetic structures that are formed by $\pm m$ and $\pm m^\prime$ scalar magnetic moments. Considering the magnetic alignments along the $\alpha$ direction yields four vector magnetic structures [see Figs.~\ref{fig:reorientation}(b) and (c)], where the vector magnetic moments are represented by $\pm m~\mathbf{e}_\alpha$ and $\pm m^\prime~\mathbf{e}_\alpha$  ($\mathbf{e}_\alpha$ being a unit vector).

\linespread{1.3}
\begin{table}[h]
    \caption{\label{tab:opertaions} The SPG, MPG and their symmetry operations for Ba$_2$CoGe$_2$O$_7$ associated 
    with the $\mathbf{e}_x$ orientation vector (see text for the definitions of the symbols).}
\begin{ruledtabular}
\footnotesize
    \begin{tabular}{c c c c c}
     &\multicolumn{4}{c}{Symmetry operations}\\
    \hline   

    \multirow{4}{*}{SPG}&  
    $\{C_{\phi,x}||\mathfrak{1}\}$  & 
    $\{C_{\phi,x}||\mathfrak{2}_z\}$ &
    $\{C_{\pi,\perp x}||\mathfrak{2}_x\}$ & 
    ~$\{C_{\pi,\perp x}||\mathfrak{2}_y\}$ \\
    ~&$\{C_{\pi,\perp x}T||\mathfrak{1}\}$  & 
    $\{C_{\pi,\perp x}T||\mathfrak{2}_z\}$ &
    $\{C_{\phi,x} T||\mathfrak{2}_x\}$ &  
    $\{C_{\phi,x} T||\mathfrak{2}_y\}$ \\
    ~&$\{C_{\phi,x}||\mathfrak{\bar{4}}_z^+\}$  & 
    $\{C_{\phi,x}||\mathfrak{\bar{4}}_z^-\}$ &
    $\{C_{\pi,\perp x}||\mathfrak{m}_{x\bar{y}}\}$ & 
    $\{C_{\pi,\perp x}||\mathfrak{m}_{xy}\}$  \\
    ~&$\{C_{\pi,\perp x}T||\mathfrak{\bar{4}}_z^+\}$  & 
    $\{C_{\pi,\perp x}T||\mathfrak{\bar{4}}_z^-\}$  &
    $\{C_{\phi,x} T||\mathfrak{m}_{x\bar{y}}\}$ & 
    $\{C_{\phi,x} T||\mathfrak{m}_{xy}\}$ 
    \\ 
    \hline
    
    MPG & $\{\mathfrak{1}||\mathfrak{1}\}$  &
          $\{\mathfrak{2}_xT||\mathfrak{2}_x\}$  &
          $\{\mathfrak{2}_y||\mathfrak{2}_y\}$  &
          $\{\mathfrak{2}_zT||\mathfrak{2}_z\}$  \\

    \end{tabular}
\end{ruledtabular}
\end{table}

In the presence of SOI, the orientation vector of a collinear magnet determines its magnetic point group (MPG) and such an MPG governs its physical properties such as polarization and magnetization (see e.g., Refs.~\cite{noncollinear1,noncollinear3,vsmejkal2022anomalous}). We take Ba$_2$CoGe$_2$O$_7$ as an example to illustrate this point. Previous works~\cite{BCGOcrystal,BCGEmagstr,BCGOmag100,BCGEaplswChong,BCGOphysica2003,BCGOmae,Ba2CoGe2O7,Ba2CoGe2O72,Ba2MnGe2O7,BCGEdiffmagstr,BCGEPicozzi,BCGOpolarline} suggest that Ba$_2$CoGe$_2$O$_7$ has a $P\bar{4}2_1 m$ space group and a scalar magnetic structure schematized in Fig.~\ref{fig:reorientation}(d). Working with $\mathbf{e}_x$, $\mathbf{e}_{xy}$ and $\mathbf{e}_z$ orientation vectors yields three vector magnetic structures [see Figs.~\ref{fig:reorientation}(e) and (f)]. According to Refs.~\cite{Ba2CoGe2O7,Ba2CoGe2O72,BCGEmagstr,BCGEdiffmagstr,BCGOmag100,findsymwb}, the MPGs of Ba$_2$CoGe$_2$O$_7$ associated with $\mathbf{e}_x$, $\mathbf{e}_{xy}$ and $\mathbf{e}_z$ orientation vectors are $2_x^\prime 2_y2_z^\prime$, $m_{xy}^\prime m_{x \bar{y}}2_z^\prime$, and $\bar42m.1$, respectively~\footnote{The subscripts (if any) in the MPG symbols label the directions of the corresponding symmetry elements. For instance, the $x$ subscript in $2^\prime_x$ means that the twofold rotation is along the $x$ direction. The details can be found in caption of Table~\ref{tab:SPGs}.}. Here, $\bar42m.1$ enables neither magnetization nor polarization, $2_x^\prime 2_y2_z^\prime$ enables a magnetization along the $y$ direction, and $m_{xy}^\prime m_{x\bar{y}}2_z^\prime$ enables a magnetization along the $x\bar{y}$ direction as well as a polarization along the $z$ direction. \\

\noindent
\textit{Spin orientation driven phenomena.} We now explore the spin orientation driven phenomena in collinear magnets. We outline our derivations and analyses by working with our aforementioned Ba$_2$CoGe$_2$O$_7$. The point group symmetry of Ba$_2$CoGe$_2$O$_7$ is described by the $^{1}\bar{4}^{\bar{1}}2^{\bar{1}}m^{\infty m}1$ spin point group (SPG) when neglecting the coupling between the spin and spatial spaces. As shown in Table~\ref{tab:opertaions}, the $^{1}\bar{4}^{\bar{1}}2^{\bar{1}}m^{\infty m}1$ SPG is an infinite group that contains 16 types of symmetry operations. These operations have the form of $\{O_s||O_l\}$ where $O_s$ acting on spin space and $O_l$ acting on spatial space are independent with each other. The $O_l$ operations include $\mathfrak{1}$, $\mathfrak{2}_\alpha$, $\mathfrak{m}_\beta$ and $\mathfrak{\bar{4}}^{\pm}_z$, which represent identity operation, twofold rotation along the $\alpha$ direction ($\alpha=x,y,z$), mirror plane perpendicular to the $\beta$ direction ($\beta=xy,x\bar{y}$), and fourfold rotoinversion along the $z$ direction, respectively. In case of the $\mathbf{e}_x$ orientation vector, the $O_s$ operations are written as $C_{\phi,x}$, $C_{\pi,\perp x}$, $C_{\phi,x}T$ and $C_{\pi,\perp x}T$: $C_{\phi,x}$ denotes the rotation of $\phi$ angle ($0<\phi \le 2\pi$) along the $x$ direction, $C_{\pi,\perp x}$ represents the rotation of $\pi$ angle along an axis perpendicular to the $x$ direction, and $T$ is a time-reversal operation. 

In reality, the presence of SOI causes the inevitable coupling between spin and spatial spaces. This will lower the $^{1}\bar{4}^{\bar{1}}2^{\bar{1}}m^{\infty m}1$ SPG symmetry to an MPG symmetry. According to Table~\ref{tab:opertaions}, the magnetic alignment along the $\mathbf{e}_x$ orientation involves the $\{\mathfrak{1}||\mathfrak{1}\}$, $\{\mathfrak{2}_y||\mathfrak{2}_y\}$, $\{\mathfrak{2}_zT||\mathfrak{2}_z\}$, and $\{\mathfrak{2}_xT||\mathfrak{2}_x\}$ symmetry operations. Such operations are essentially $\mathfrak{1}$, $\mathfrak{2}_y$, $\mathfrak{2}_z^\prime$, and $\mathfrak{2}_x^\prime$ MPG symmetry operations, which collectively form an MPG of $2_x^\prime2_y2_z^\prime$. We also consider the cases with magnetic alignments characterized by $\mathbf{e}_{xy}$ and $\mathbf{e}_z$ orientation vectors (the SPG being $^{1}\bar{4}^{\bar{1}}2^{\bar{1}}m^{\infty m}1$ as well). The MPGs associated with $\mathbf{e}_{xy}$ and $\mathbf{e}_z$ are $m_{xy}^\prime m_{x \bar{y}}2_z^\prime$ and $\bar{4}2m.1$, respectively. The details of our symmetry analysis regarding Ba$_2$CoGe$_2$O$_7$ are shown in Section I of the Supplementary Material (SM).

Our aforementioned analyses together with Refs.~\cite{Libor2022Beyond,Libor2022Emerging,lqh2022} deliver three pieces of information. First, a collinear magnet in the regime of decoupled spin and spatial spaces can be ascribed to a collinear SPG regardless of its orientation vectors. Second, the point group symmetry of a collinear magnet in the regime of coupled spin and spatial spaces is described by an MPG. Third, the MPG is a subgroup of the SPG, and the operations of the MPG should be compatible with the orientation vector. These three aspects motivate us to explore the spin orientation driven phenomena by working with collinear SPGs. In this regard, we map each $G_{sp}$ SPG to several $G^{\alpha}_{sp}, G^{\beta}_{sp}, \cdots$ MPGs that are associated with $\mathbf{e}_\alpha,\mathbf{e}_\beta,\cdots$, where $\mathbf{e}_\alpha,\mathbf{e}_\beta,\cdots$ are selected as orientation vectors along high symmetric $\alpha,\beta,\cdots$ crystallographic directions. We then examine whether $G^{\alpha}_{sp}, G^{\beta}_{sp}, \cdots$ are symmetrically compatible with our interested phenomena. In Tables S3--S9 of the SM, we analyze the 122 collinear SPGs with respect to spin reorientation driven polarization and magnetization in collinear magnets. We concentrate on spin orientation driven polarization, and list in Table~\ref{tab:SPGs} the nonpolar SPGs that host such a phenomenon.

Before finishing this section, we point out that several polar SPGs may host spin orientation driven polarization as well. For instance, the $^{1}3^{\infty m}1$, $^{1}3^{\infty /mm}1$, $^{1}3^{1}m^{\infty m}1$, $^{1}3^{\bar1}m^{\infty m}1$, and $^{1}3^{1}m^{\infty /mm}1$ SPGs enable a polarization along the $z$ direction; the presence of SOI together with the $\mathbf{e}_x$ or $\mathbf{e}_y$ orientation vector yields MPGs that enable additional polarization components along the $x$ and/or $y$ directions (see e.g., Table S7 of the SM). These polar SPGs are not shown in Table~\ref{tab:SPGs}.\\

\begin{table}[htbp]
\caption{\label{tab:SPGs} Nonpolar SPGs hosting spin orientation driven polarization. The second, third, and fourth columns list the MPGs associated with special orientation vectors. The subscripts are marked in the symmetry elements of an MPG to label the direction of the rotation/rotoinversion axis or the normal of the mirror plane, if the MPG does not belong to the standard UNI~\cite{msg,MPOINT} notation; $u$, $v$, and $w$ represent $xy$, $x\bar{y}$, and $xyz$ directions, respectively. The $x$, $y$, $z$, $u$, $v$ or $w$ symbol in the parentheses labels the direction of the polarization associated with the MPG.}
\renewcommand{\arraystretch}{1.3}
\begin{ruledtabular}
    \centering
    \footnotesize
    \begin{tabular}{l l l l  }   
 SPGs & $\mathbf{e}_x$  & $\mathbf{e}_z$  & $\mathbf{e}_{xy}$  \\
\hline

  $^1\bar{4}^{\infty m}1$& $2_z^\prime$~($z$) & $\bar4.1$ & $2_z^\prime$~($z$) \\
  $^{\bar1}\bar{4}^{\infty m}1$ & $2_z^\prime$~($z$)  & $\bar4^\prime$  & $2_z^\prime$~($z$)  \\
  $ ^1\bar{4}^{\infty/m m}1$& $2_z.1^\prime$~($z$) & $\bar4.1^\prime$ & $2_z.1^\prime$~($z$) \\

   $^{1}\bar{4}^{1}2^{1}m^{\infty m}1$& $2_x2_y^\prime2_z^\prime$ &  $\bar42^\prime m^\prime$ & $m_{u}m_{v}^\prime 2_z^\prime$~($z$)\\
  $^{\bar1}\bar{4}^{\bar1}2^{1}m^{\infty m}1$ & $2_x^\prime 2_y2_z^\prime$  &$\bar4^\prime 2m^\prime $ & $m_{u}m_{v}^\prime 2_z^\prime$~($z$) \\
 $^{\bar1}\bar{4}^{1}2^{\bar1}m^{\infty m}1$ &  $2_x2_y^\prime2_z^\prime $ &  $\bar4^\prime 2^\prime m$ & $m_{u}^\prime m_{v}2_z^\prime$~($z$) \\
 $^{1}\bar{4}^{\bar1}2^{\bar1}m^{\infty m}1$ & $2_x^\prime 2_y2_z^\prime$  &  $\bar42m.1$ & $m_{u}^\prime m_{v}2_z^\prime$~($z$)\\
  $^{1}\bar{4}^{1}2^{1}m^{\infty/m m}1$ & $222.1^\prime$ &  $\bar42m.1^\prime$ & $m_{u}m_{v}2_z.1^\prime$~($z$) \\
\hline 
\hline

 SPGs & $\mathbf{e}_x$ & $\mathbf{e}_y$& $\mathbf{e}_z$   \\
\hline
  $^{1}3^{1}2^{\infty m}1$& $2_x.1$~($x$) & $2_x^\prime$~($x$) & $32^\prime$   \\
  $^{1}3^{\bar1}2^{\infty m}1$ & $2_x^\prime$~($x$) & $2_x.1$~($x$) & $32.1$  \\
 $^{1}3^{1}2^{\infty/m m}1$& $2_x.1^\prime~$($x$) & $2_x.1^\prime~$($x$) & $32.1^\prime$   \\
   
   $^{1}\bar{6}^{\infty m}1$& $m_z^\prime$~($x$, $y$) & $m_z^\prime$~($x$, $y$) & $\bar6.1$\\
 $^{\bar1}\bar{6}^{\infty m}1$ & $m_z.1$~($x$, $y$) & $m_z.1$~($x$, $y$) & $\bar6^\prime$ \\
 $^{1}\bar{6}^{\infty/m m}1$& $m_z.1^\prime$~($x$, $y$) & $m_z.1^\prime$~($x$, $y$) & $\bar6.1^\prime$   \\

  $^{1}\bar{6}^{1}m^{1}2^{\infty m}1$& $m_xm_z^\prime 2_y^\prime$~($y$) & $m_x^\prime m_z^\prime 2_y$~($y$) & $\bar6m^\prime 2^\prime$   \\
 $^{\bar1}\bar{6}^{\bar1}m^{1}2^{\infty m}1$ & $m_x^\prime m_z2_y^\prime$~($y$) & $m_xm_z2_y.1$~($y$) &  $\bar6^\prime m2^\prime$ \\
 $^{\bar1}\bar{6}^{1}m^{\bar1}2^{\infty m}1$ & $m_xm_z2_y.1$~($y$) & $m_x^\prime m_z2_y^\prime$~($y$) &  $\bar6^\prime m^\prime2$ \\
 $^{1}\bar{6}^{\bar1}m^{\bar1}2^{\infty m}1$ & $m_x^\prime m_z^\prime 2_y$~($y$) & $m_xm_z^\prime 2_y^\prime$~($y$) & $\bar6m2.1$ \\
 $^{1}\bar{6}^{1}m^{1}2^{\infty/m m}1$ & $m_xm_z2_y.1'$~($y$) & $m_xm_z2_y.1^\prime$~($y$) & $\bar6m2.1^\prime$   \\
\hline 
\hline

SPGs  & $\mathbf{e}_z$ & $\mathbf{e}_{xy}$ & $\mathbf{e}_{xyz}$ \\
\hline
 
 $^{1}2^{1}3^{\infty m}1$ & $2^\prime 2^\prime2$ & $2_z^\prime$~($z$) & $3_{w}.1$~($w$)\\
 $^{1}2^{1}3^{\infty/m m}1$ & $222.1^\prime$& $2_z.1^\prime$~($z$) & $3_{w}.1^\prime$~($w$)\\

 $^{1}\bar{4}^{1}3^{1}m^{\infty m}1$ & $\bar42^\prime m^\prime$ & $m_{u}m_{v}^\prime 2_z^\prime$~($z$) & 
 $3_{w}m_{v}^\prime$~($w$)\\
 $^{\bar{1}}\bar{4}^{1}3^{\bar{1}}m^{\infty m}1$ & $\bar4^\prime 2^\prime m$& $m_{u}^\prime m_{v}2_z^\prime$~($z$) & 
 $3_{w}m_{v}.1$~($w$)\\
 $^{1}\bar{4}^{1}3^{1}m^{\infty/m m}1$ & $\bar42m.1^\prime$ & $m_{u}m_{v}2_z.1^\prime$~($z$) & $3_{w}m_{v}.1^\prime$~($w$)\\

    \end{tabular}
\end{ruledtabular}
\end{table}

\noindent
\textit{Spin orientation driven polarization in collinear magnets.} Spin orientation driven phenomena are commonly found in collinear magnets (see e.g., Refs.~\cite{noncollinear1,noncollinear3,vsmejkal2022anomalous}). In the following, we concentrate on spin orientation driven polarization in type-II multiferroics and related materials. Previous experiments identify Ba$_2$CoGe$_2$O$_7$~\cite{Ba2CoGe2O7,Ba2CoGe2O72,Ba2MnGe2O7,BCGEdiffmagstr,BCGEPicozzi,BCGOpolarline}, Sr$_2$CoSi$_2$O$_7$~\cite{Sr2CoSi2O7,SCSOmag}, Sr$_2$CoGe$_2$O$_7$~\cite{Sr2CoMnGeO,SrCoGeOpolar} and Ca$_2$CoSi$_2$O$_7$~\cite{CCSOapljapanmag,CCSOnc} as type-II multiferroics whose magnetic structures resemble those for Ba$_2$CoGe$_2$O$_7$ [see Fig.~\ref{fig:reorientation}(f)]. The SPG of these materials is $^{1}\bar{4}^{\bar{1}}2^{\bar{1}}m^{\infty m}1$, and it is shown that the $\mathbf{e}_{xy}$-oriented magnetic alignment yields an MPG of $m_{xy}^\prime m_{x \bar{y}}2_z^\prime$ and a polarization along the $z$ direction~\cite{Ba2CoGe2O7,Ba2CoGe2O72,Ba2MnGe2O7,BCGEdiffmagstr,BCGEPicozzi,BCGOpolarline,Sr2CoSi2O7,SCSOmag,Sr2CoMnGeO,SrCoGeOpolar,CCSOapljapanmag,CCSOnc}. Other experimentally confirmed representative type-II multiferroics include (i) Ba$_2$MnGe$_2$O$_7$~\cite{Ba2MnGe2O7,Ba2MnGe2O7prb} and Ba$_2$FeSi$_2$O$_7$~\cite{Ba2FeSi2O7,BFSO2} with the $^{1}\bar{4}^{1}2^{1}m^{\infty/m m}1$ SPG, the $\mathbf{e}_{xy}$ orientation vector and the $m_{xy}m_{x \bar{y}}2_z.1^\prime$ MPG, (ii) $A$Mn$_3$Cr$_4$O$_{12}$~($A$ = La, Sm, Tb)~\cite{LaMnCrOlong,LMCOHJZ,LMCOHJX,SmMnCrO,TbMnCrO,TbMnCrOtheory} with the $^{1}2^{1}3^{\infty/m m}1$ SPG, the $\mathbf{e}_{xyz}$ orientation vector and the $3_{xyz}.1^\prime$ MPG, and (iii) Cu$_2$OSeO$_3$~\cite{Cu2OSeO3science,Cu2OSeO3prb,Cu2OSeO3} with the $^{1}2^{1}3^{\infty m}1$ SPG, the $\mathbf{e}_{xyz}$ orientation vector and the $3_{xyz}.1$ MPG. Furthermore, first-principles simulations~\cite{MgFeN,OsX2,VSVSe,VSeTeRuBr,GdI2} predict a sequence of type-II multiferroics with spin orientation driven polarization. Particularly, Ref.~\cite{MgFeN} classifies layer groups with respect to spin orientation driven polarization in two-dimensional altermagnets with two magnetic sublattices and predicts monolayer MgFe$_2$N$_2$ as a candidate exhibiting such a phenomenon; Ref.~\cite{OsX2} predicts OsCl$_2$, OsBr$_2$ and OsI$_2$ as room-temperature type-II multiferroics with polarization larger than 1 $\mu$C/cm$^2$. In Tables S10 and S11 of the SM, we list the SPGs, the orientation vectors and the MPGs for various type-II multiferroics. The mechanisms for the spin-driven polarization in these multiferroic materials can be interpreted by our Table~\ref{tab:SPGs}.

\begin{figure}[htbp]
\includegraphics[width=1.0\linewidth]{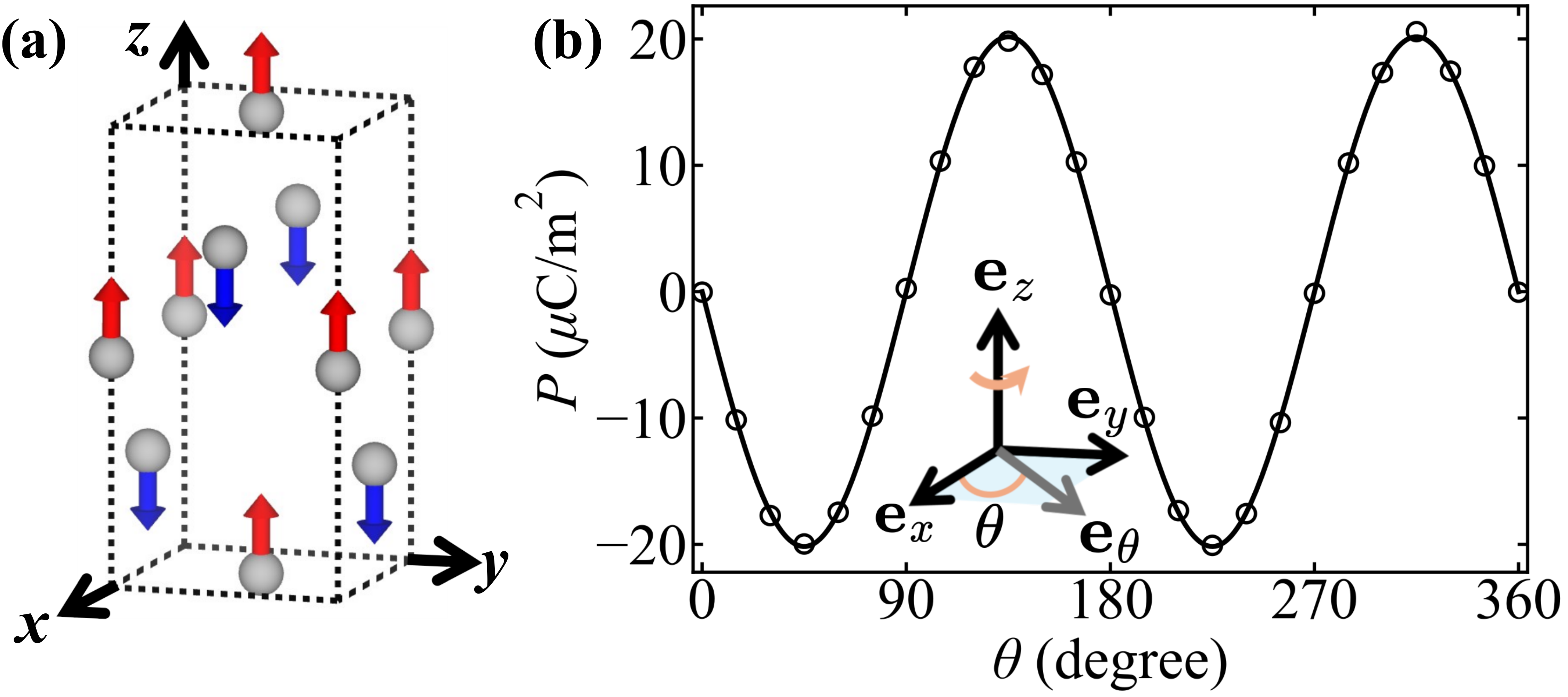} 
\caption{Spin reorientation driven polarization in CuFeS$_2$. Panel (a) sketches the ground state magnetic structure associated with a $\mathbf{e}_z$ orientation vector. The Fe ions are denoted by gray spheres, while the Cu and S ions are not displayed. The red and blue arrows represent the vector magnetic moments carried by Fe ions. Panel (b) shows the polarization of CuFeS$_2$ as a function of the $\theta$ orientation angle. As defined in the inset of panel (b), $\theta$ is the angle between $\mathbf{e}_x$ and $\mathbf{e}_\theta$, where $\mathbf{e}_\theta$ lies in the plane perpendicular to $\mathbf{e}_z$. The first-principles results are marked by unfilled circles, and the solid line is obtained by a sinusoidal function fitting (coefficient of determination being about 99.98\%).}
\label{fig:polar}
\end{figure}

On the other hand, spin-orientation-driven polarization is not limited to type-II multiferroics but also occurs in \textit{nonpolar} magnets. In this regard, we concentrate on \textit{nonpolar} collinear magnets whose polarization can be induced by reorienting their magnetic alignments (i.e., spin reorientation). By examining the MAGNDATA database~\cite{magndata} and Refs.~\cite{BMSO,NaMnFeF6,CsCoF,CuFeS,CuFeStemperature,LiFeCro,ZnFe2O4,TbFeBO}, we identify a sequence of oxides, fluorides and sulfides that exhibit spin reorientation driven polarization and summarize them in Table S12 of the SM. Of particular interest is the room-temperature antiferromagnetic CuFeS$_2$ with a N\'eel temperature of 823 K~\cite{CuFeStemperature} and an SPG of $^{1}\bar{4}^{\bar1}2^{\bar1}m^{\infty m}1$. Figure~\ref{fig:polar}(a) sketches the magnetic structure of CuFeS$_2$ with an orientation vector $\mathbf{e}_z$. Such a magnetic alignment results in a nonpolar $\bar42m.1$ MPG (see Table~\ref{tab:SPGs}). The $\mathbf{e}_x$ and $\mathbf{e}_{xy}$ orientation vectors are associated with nonpolar $2^\prime_x 2_y 2^\prime_z$ MPG and polar $m^\prime_{xy} m_{x\bar{y}} 2^\prime_z$ MPG, respectively. The $m^\prime_{xy} m_{x\bar{y}} 2^\prime_z$  MPG enables a polarization along the $z$ direction. We use first-principles simulations to explore the spin reorientation driven polarization in CuFeS$_2$, focusing on a variety of $\mathbf{e}_\theta$ orientation vectors [see the inset of Fig.~\ref{fig:polar}(b)]. As shown in Figure~\ref{fig:polar}(b), the polarization of CuFeS$_2$ behaves as a periodic sinusoidal function of the $\theta$ orientation angle with a period of $180\degree$, where $\theta=45\degree$ and $\theta=225\degree$ correspond to $+\mathbf{e}_{xy}$ and $-\mathbf{e}_{xy}$ orientation vectors, respectively. Rotating the $\pm\mathbf{e}_{xy}$ orientation vectors by $90\degree$ yields $\theta=135\degree$ and $\theta=315\degree$ magnetic alignments with a reversed polarization. Similar phenomena have also been predicted to occur in CaFeO$_2$ and MgFeO$_2$, whose polarizations and orientation angles showcase sinusoidal relationship~\cite{CaFeOpicozzi}.\\

\noindent
\textit{Summary and outlook.} To summarize, we have established a symmetry-based theory on spin orientation driven phenomena in collinear magnets. The application of our theory yields the classification of collinear SPGs with respect to spin orientation driven polarization. Such a classification not only interprets the origin of polarization in a variety of type-II multiferroics, but also is helpful to identify nonpolar collinear magnetic materials with spin reorientation induced polarization. Regarding the latter, we find a sequence of materials that are originally nonpolar but can be polar by reorienting their magnetic alignments. Further, we use first-principles simulations to predict that nonpolar antiferromagnetic CuFeS$_2$ is a candidate with spin reorientation driven polarization. In practice, the orientation vectors of collinear ferromagnets can be directly switched by magnetic field. As for collinear antiferromagnets, the orientation vectors can be reoriented magnetically via interfacial exchange coupling with an adjacent magnetic substrate~\cite{park2011spinNM} or electrically thanks to an asymmetric spin torque~\cite{SpinTorqueshao,hanxiufeng2024electrical}. These means can definitely be employed to generate our aforementioned spin reorientation driven polarization in nonpolar collinear magnets. As an outlook, our work provides guidelines for discovering type-II multiferroics and related materials with spin orientation driven phenomena, which will benefit the development of spintronics.\\

\noindent
\textit{Acknowledgements.---} The authors acknowledge the support from the Advanced Materials-National Science and Technology Major Project (Grant No.~2024ZD0606900) and the support from the National Natural Science Foundation of China (Grants No.~12274174, No.~52288102, No.~52090024, and No.~12034009). Y.Z. thanks the support from high-performance computing center of Jilin University. H.J.Z. thanks the support from ``Xiaomi YoungScholar'' Project.

\end{document}